\documentclass[conference, a4paper]{IEEEtran}
\IEEEoverridecommandlockouts                             
\usepackage{units}
\usepackage{graphicx}
\usepackage{multirow, makecell}
\usepackage{amsmath,amssymb,latexsym}
\usepackage{rotating}
\usepackage{lettrine}
\usepackage{textcomp}

\usepackage{hyperref}
\usepackage{lipsum}% 
\usepackage[pscoord]{eso-pic}
\newcommand{\placetextbox}[3]{% 
  \setbox0=\hbox{#3}
  \AddToShipoutPictureFG*{
    \put(\LenToUnit{#1\paperwidth},\LenToUnit{#2\paperheight}){\vtop{{\null}\makebox[0pt][c]{#3}}}%
  }%
}%

\usepackage{bm}

\title{Packet Steering Mechanisms for MLO in Wi-Fi 7
\thanks{This work has been partially funded by SoBigData.it ``SoBigData.it receives funding from European Union – NextGenerationEU – National Recovery and Resilience Plan (Piano Nazionale di Ripresa e Resilienza, PNRR) – Project: “SoBigData.it – Strengthening the Italian RI for Social Mining and Big Data Analytics” – Prot. IR0000013 – Avviso n. 3264 del 28/12/2021.''}}

\author{
    \IEEEauthorblockN{
    Gianluca Cena\IEEEauthorrefmark{1},
    Matteo Rosani\IEEEauthorrefmark{1}\IEEEauthorrefmark{2},
    Stefano Scanzio\IEEEauthorrefmark{1}
    }  
    \IEEEauthorblockA{\IEEEauthorrefmark{1}National Research Council of Italy (CNR--IEIIT), Italy --- \IEEEauthorrefmark{2}Politecnico di Torino, Italy.}    
    Email:  \{gianluca.cena, stefano.scanzio\}@cnr.it, \{matteo.rosani\}@polito.it
}

\begin{document}
\placetextbox{0.5}{1}{This is the author's version of an article that has been published.}
\placetextbox{0.5}{0.985}{Changes were made to this version by the publisher prior to publication.}
\placetextbox{0.5}{0.97}{The final version of record is available at \href{https://doi.org/10.1109/ETFA61755.2024.10710726}{https://doi.org/10.1109/ETFA61755.2024.10710726}}%
\placetextbox{0.5}{0.05}{Copyright (c) 2024 IEEE. Personal use is permitted.}
\placetextbox{0.5}{0.035}{For any other purposes, permission must be obtained from the IEEE by emailing pubs-permissions@ieee.org.}%

\maketitle
\thispagestyle{empty}
\pagestyle{empty}

%%%%%%%%%%%%%%%%%%%%%%%%%%%%%%%%%%%%%%%%%%%%%%%%%%%%%%%%%%%%%%%%%%%%%%%%%%%%%%%%
\begin{abstract}
Besides extremely high throughput, \mbox{Wi-Fi} 7 is also aimed at providing users a more deterministic behavior, characterized by shorter average latency and smaller jitters.
A key mechanism to achieve this is multi-link operation, which brings simultaneous multi-band communication to client stations as well.

In this paper, traffic steering policies are briefly reviewed and grouped into general classes, each one with its advantages and limitations.
A basic mechanism for supporting dynamic steering is then described, which is simple enough 
to allow implementation in real \mbox{Wi-Fi} chipsets but highly flexible at the same time.
Its operation can be driven by the host on a per-packet basis, and this permits to optimize spectrum usage depending on the requirements of applications and the traffic pattern they generate.
\end{abstract}

%%%%%%%%%%%%%%%%%%%%%%%%%%%%%%%%%%%%%%%%%%%%%%%%%%%%%%%%%%%%%%%%%%%%%%%%%%%%%%%%

\section{Introduction}
\label{sec:intro}

\begin{figure}[b]
    \vspace{-0.4cm}
    \centering
    \includegraphics[width=1.0\columnwidth]{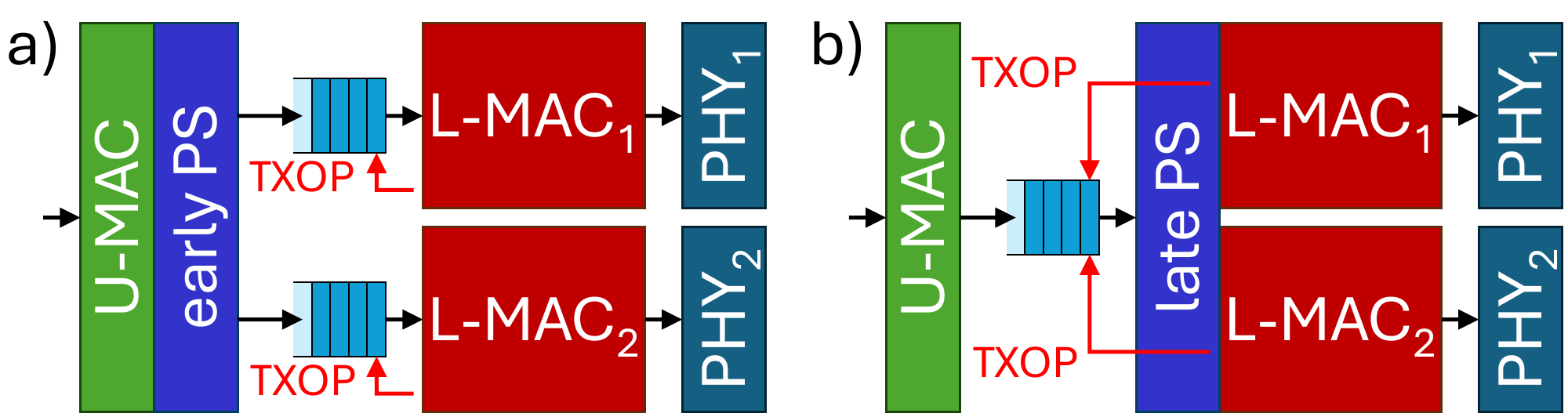}
    \caption{Conceptual MLD architecture (early and late packet steering).}
    \label{fig:MLO}
\end{figure}

One of the most peculiar (and potentially groundbreaking) 
novelties implemented by the media access control (MAC) layer of \mbox{Wi-Fi} 7 is Multi-Link Operation (MLO)
\cite{2022-IEEE-WCOM}.
As depicted in Fig.~\ref{fig:MLO}, 
a Multi-Link Device (MLD) consists in two (or more) affiliated STAs, denoted \mbox{L-MAC}s, 
tuned on different channels (typically chosen in different bands, for instance $2.4$, $5$, and $\unit[6]{GHz}$),
which take care of managing access to the shared spectrum and frame transmission on the related PHYs.

Every time a transmission request is made for a packet by the user of the data-link layer (typically IP), a specific entity in the MLD known as \mbox{U-MAC} takes care of deciding which one among the available \mbox{L-MAC}s will be used for transmitting it on air \cite{2022-IEEE-WCOMLET}.
This function is sometimes denoted \textit{packet steering}.
Decisions are made starting from the available information about the Quality of Service (QoS) a certain packet (or the flow it belongs to) expects to receive from the network.
To maximize overall performance, the capabilities and current state of all L-MACs and the related channels can be also considered.
For example, the wider bandwidth available in the $5$ and $\unit[6]{GHz}$ bands makes them more suitable for 
multimedia traffic (video and voice), which needs high throughput and low latency.
Conversely, the $\unit[2.4]{GHz}$ band typically offers 
wider coverage and full compatibility with legacy devices.

Besides the specific channel the selected \mbox{L-MAC} is tuned on, also the access category (AC) used to send the packet (voice, video, best effort, and background) will impact on the actual QoS it receives.
Parameters of the enhanced distributed channel access (EDCA) \cite{IEEE802-11-2021}, defined on a per-AC basis, include the minimum and maximum sizes of the contention window (CW) and the arbitration inter-frame spacing (AIFS), which determine the access priority when different STAs are trying to access the channel at the same time, as well as the duration of transmission opportunities (TXOP).

In the following, different kinds of packet steering mechanisms are briefly analyzed, highlighting their advantages and drawbacks.
Then, some approaches are proposed that focus on flexibility and practical feasibility of their implementations.
The paper is structured as follows:
Section~\ref{sec:arch} describes a sample reference architecture for the network controller,
Section~\ref{sec:steer} provides a tentative taxonomy of packet steering mechanisms, 
whereas Section~\ref{sec:prop} introduces our simple proposal for supporting dynamic channel selection driven by the host.
Finally, in Section~\ref{sec:concl} some conclusions are drawn.

\section{Reference Architecture}
\label{sec:arch}
To assess practical feasibility of MLO packets steering solutions, the architecture of real chipsets has to be considered.
In the following we will refer to the Atheros AR9344 System-on-a-Chip, a highly-integrated platform that supports IEEE 802.11n (\mbox{Wi-Fi} 4) on the $2.4$ and $\unit[5]{GHz}$ bands and permits the implementation of  WLAN devices \cite{AR9344}.
Although it is not recent, most of its functions at the MAC level are basically the same as \mbox{Wi-Fi} 5 (that relies on extended but very similar operating principles).
Concerning \mbox{Wi-Fi} 6, in this preliminary study we will not consider orthogonal frequency division multiple access (OFDMA), multi-user multiple-input multiple-output (MU-MIMO), and not even trigger frames, and hence what we propose can be applied to this case as well.
Suitable extensions to tackle these features will be the subject of future work.
Exact details of the AR9344 platform are mostly irrelevant: it just as an example to avoid re-inventing the wheel and re-defining abstract terms to denote well-known concepts.

\begin{figure}[b]
    \vspace{-0.3cm}
    \centering
    \includegraphics[width=0.9\columnwidth]{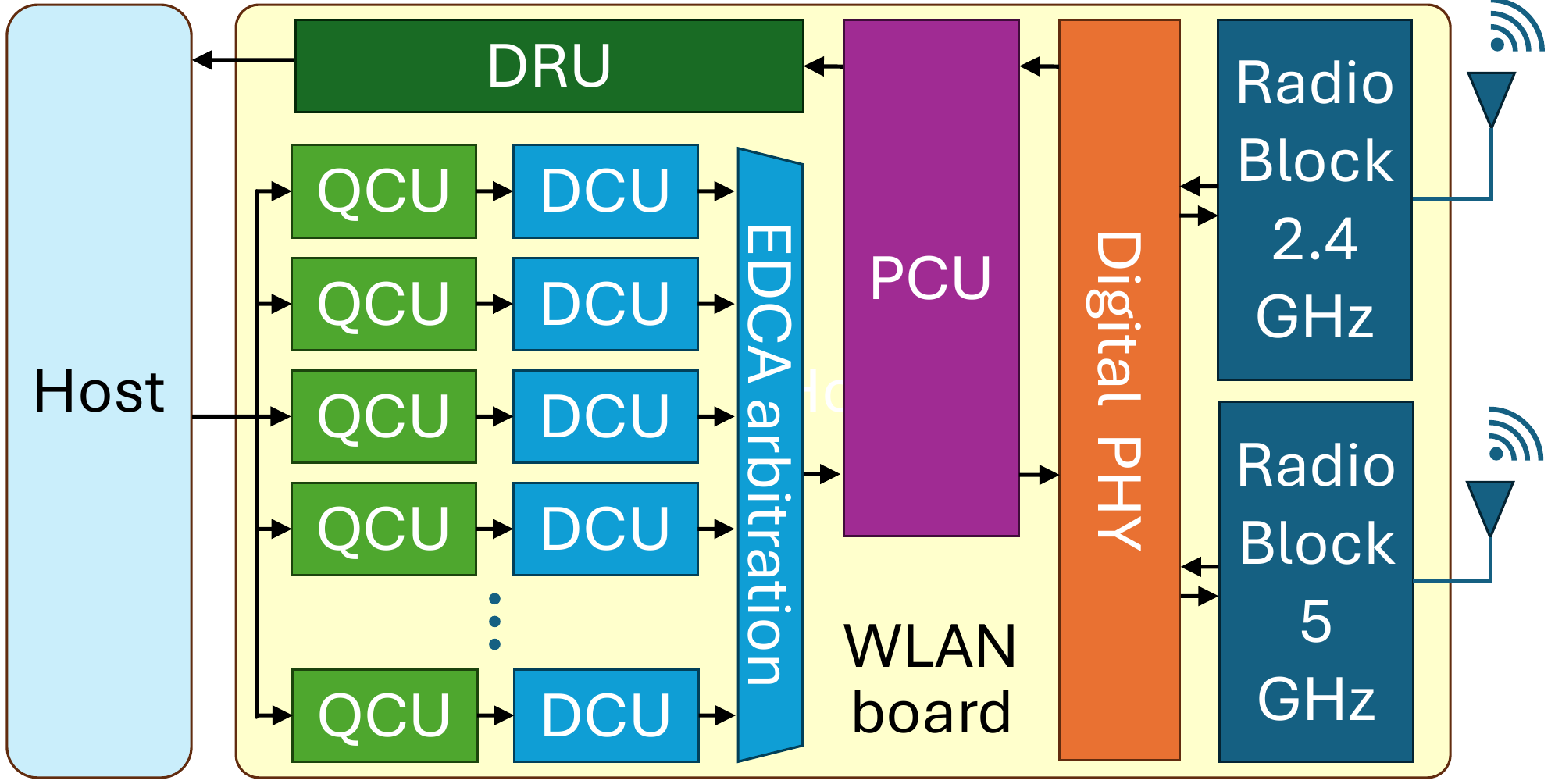}
    \caption{Real WLAN adapter architecture with QoS support (ACs).}
    \label{fig:AR}
\end{figure}

As shown in Fig.~\ref{fig:AR}, the architecture of a \mbox{Wi-Fi} controller is made up of several parts.
The \textit{transmitting} path consists of a number of parallel units, each one made up of a pair of blocks, that is, a 
queue control unit (QCU) and a distributed coordination function (DCF) control unit (DCU).
Every QCU-DCU pair (in the considered architecture, ten are available) manages a specific transmission buffer.
For example, a separate buffer is employed for every AC, plus one for beacons and another for beacon-gated frames.
In particular, the QCU fetches descriptors for the packets to be sent from the host memory using direct memory access (DMA) and feeds them to the related DCU.
In turn, the DCU manages media access according to the EDCA rules (including exponential backoff and virtual collisions).
As soon as one of the DCUs gains access to the channel, one or more queued frames are moved to the protocol control unit (PCU), which performs the required functions (e.g., encryption) before passing the frame to the baseband logic (digital PHY).

The \textit{receiving} path starts instead in the digital PHY, which passes the incoming bit stream to the PCU to get back a full-formed frame (decryption and frame check sequence verification are performed here).
The frame is then copied in the host memory by the DMA receive unit (DRU).

A distinct radio block is foreseen for every band ($2.4$ and $\unit[5]{GHz}$).
They are placed beneath the digital PHY and take care of transmission and reception on air.
These latter blocks are not relevant for our discussion, since transmission optimization only involves the MAC layer on the sender side.

\section{Packet Steering for MLO}
\label{sec:steer}
Packet steering mechanisms are aimed at deciding on which channel any given packet must be transmitted, 
by selecting one of the available \mbox{L-MAC}s in such a way to optimize communication.
As reported in Table~\ref{tab:TAX}, several kinds of approaches can be devised to this purpose, which mainly differ for the instant when the steering decision is made (on packet transmission request, TXreq, or upon TXOP acquisition) and the information on which it is based (extracted from the message or acquired from the context).

\begin{table}
    \caption{Simple taxonomy of packet steering mechanisms}
    \label{tab:TAX}
    \tabcolsep=0.11cm 
    \centering
    \begin{tabular}{cll}
         Type               & When         & How \\
         \hline
         \textit{Early}     & on\! TXreq   & Static:  info in protocol headers (RA, ToS, Port) \\
                  &              & Dynamic: info from status registers and statistics \\
         \hline
         \textit{Late}      & on\! TXOP       & FIFO: enqueued packets are served in strict order  \\
                  &              & Any:  enqueued packets are filtered (hardly feasible) \\
         \hline
         \textit{Split}     & on\! TXreq   & Channel bitmap is preliminary configured + \\
         (CRS)              & on\! TXOP    & Packets for which channel is enabled are orderly served \\
    \end{tabular}
    \vspace{-0.5cm}
\end{table}

\subsection{Early Steering (TXreq)}
In this case, every \mbox{L-MAC} has its own separate transmission queue and the task of the \mbox{U-MAC} is to select, for every newly generated packet (that is, when the related transmission request is issued to the data-link layer), where it must be enqueued.
Early steering is typically performed in software by the host (see Fig.~\ref{fig:MLO}.a).
The simplest approach only considers \textit{static} information embedded in the frame, e.g., receiver MAC address (RA) and AC.
However, a cross-layer design may foresee that the packet is inspected to retrieve upper-layer information, e.g., the type of service (ToS) field in the IPv4 header, which is now exploited by differentiated services (DiffServ), or the TCP/UDP port (to determine the application layer protocol, e.g., HTTP, DNS, RTP).

The biggest drawback of static steering is that, it does not adapt to variations of the spectrum conditions,
which in real scenarios may occur suddenly and unexpectedly.
For example, a narrowband jamming signal (or just a nearby node temporarily transmitting a huge amount of data over the air, as happens for large file transfers and high-definition multimedia streaming) may preclude the timely exchange of packets for data flows mapped on specific radio channels, leaving the other links of the MLD mostly untouched.

To address these issues, adaptive steering mechanisms can be envisaged \cite{2023-PEMWN}
where decisions additionally refer to the \textit{dynamic} state of the available links (which must be constantly monitored).
For example, the measured average capacity of every link, which depends on the modulation and coding scheme (MCS) in use, can be considered.
In this case, faster links must be preferably selected to properly balance traffic over channels.
The number $L$ of packets enqueued in the transmission buffer of every \mbox{L-MAC} is another relevant information they can provide to the \mbox{U-MAC} through status registers.
By using the Little's law, the average waiting time can be determined as 
$\overline W = \overline L/\lambda$, where $\lambda$ is the packet arrival rate for the link, which can be estimated either by summing the expected loads generated by applications (if they are known) or from the actual traffic by using, e.g., moving averages.
For time-aware applications, steering can be decided by checking $\overline W$ against packet deadlines.
Finally, when utmost reliability is demanded, the packet loss ratio (PLR) on the different channels could be also taken into account for decisions, privileging links where errors are experienced seldom (even if they are slower).
This is because frames can be retransmitted only a limited number of times by the MAC, after which they are silently discarded.

Making packet steering dynamically adapting to the operating conditions makes the system more predictable and robust.
Unfortunately, reaction times may be slow, because decisions taken by the \mbox{U-MAC} are based on statistics collected over time intervals in the past whose width could range from a few seconds to minutes.
This means that, if the disturbance affecting a given channel varies suddenly, the related flows may experience temporary communication outages, which can be either annoying (e.g., for voice over IP) or potentially dangerous.
The latter case concerns systems where wireless communication is used to exchange real-time process data,
e.g., when a fleet of autonomous mobile robots (AMR) is coordinating its operation over the air.
The solution proposed in IEEE 802.11be relies on traffic identifiers (TID)
and is known as TID-to-link mapping \cite{2022-IEEE-WCOM}.
Here, APs may move a TID from one link to another at runtime.

\subsection{Late Steering (TXOP)}
Intuitively, better performance could be obtained by postponing steering decisions until the time when single transmission attempts on air are about to start (and not when the packet is enqueued for transmission).
In this case, the mechanism must be managed by the network adapter.
The simplest approach of this kind foresees that all packets of the same AC are stored in the same transmission queue and, whenever one of the \mbox{L-MAC}s acquires a TXOP (i.e., when access to the related channel is obtained according to EDCA rules), they are served according to a FIFO policy (see Fig.~\ref{fig:MLO}.b).
By doing so, average latency is expected to shrink.
However, because of the strict FIFO ordering, nothing can be done in this case by the \mbox{U-MAC} to 
differentiate the QoS received by the different packets (or packet flows).

To obtain the most from steering, 
specific optimizations 
should be performed every time a TXOP is gained.
A subset of the enqueued packets can be then selected on-the-fly (frame aggregation is typically exploited) and sent on the related channel.
To improve flexibility, selection must take place according to criteria that depend on the applications' requirements and constraints, as well as on the current adapter and spectrum conditions.
Out-of-order packet arrivals on different links may be an issue for applications, and deserve further investigation (e.g., reordering buffers).
To ease reassembly, all fragments of the same frame should be sent on the same link.

Optimization algorithms could be arbitrarily complex and exploit machine learning (ML) as well.
Although not impossible, implementing them in the network adapter is not a good idea for three main reasons: 
1) the embedded $\mu$C typically lacks the required computational power;
2) adapters are typically unaware of the specific requirements of applications; and, 
3) any changes to these algorithms require the adapter firmware to be flashed, which is often inconvenient.
Nevertheless, they are typically manageable by modern CPUs, which have plenty of processing power and memory resources.

In theory, the following \textit{split} arrangement could be devised, 
which is carried out jointly by the CPU and the network adapter:
upon TXOP acquisition by a DCU, an interrupt is raised to the CPU of the host, which possesses the required knowledge (up-to-date information can be easily collected from the status registers of the network adapter) and performs the computations needed for making decisions.
The outcome is a list of descriptors, corresponding to the sequence of packets to be sent,
which is then fed to the relevant L-MAC for actual transmissions on air.
In practice, the latency involved in managing interrupts (upcalls) is too large and does not fit the tight timing constraints of EDCA (in the order of a few microseconds).
For example, the duration of the DCF interframe space (DIFS), which precedes every transmission attempt, is just $\unit[28]{\mu s}$.

\section{Simple and Flexible Steering Approach}
\label{sec:prop}
A sensible solution to traffic steering can be found by looking at rate adaptation (RA) algorithms like Minstrel \cite{minstrel, documentationminstrel}, which are widely used in commercial equipment.
In these cases the overall mechanism for selecting transmission parameters is split in two parts:
the former, very simple and fast, is executed directly by the network adapter, 
while the latter, more complex and slower, runs on the host.
A typical arrangement (available, e.g., in Atheros chipsets) foresees the presence of a number of TX series (four, in that case), each one defining the MCS to be used by a number of consecutive transmission attempts on air.
The configuration of every TX series (MCS and maximum number of attempts, but also whether or not RTS/CTS is used for channel reservation, transmission power, beamforming, etc.) is included in the packet descriptor enqueued in the transmission buffer by using a compact encoding.
When performing a transmission attempt on air (either the initial one or any of the retries), the PCU parses the descriptor to know how to do so.

One may ask how transmission parameters are determined, and who does so.
Generally speaking, reinforcement learning (RL) algorithms run on the host CPU, which measure the PLR of every link separately for the different MCSs.
Transmission attempts are performed using all the available MCSs (random \textit{exploration}), 
but the MCS offering the best QoS is selected more often to maximize overall performance (\textit{exploitation}).
Link quality is evaluated in background by a suitable software module, parallel to the operation of the network board.
It exploits the outcomes of transmission attempts (available to the driver) and requires no upcalls.
Every time a packet is enqueued for transmission, the parameters of the four TX series (also known as retry chains), determined for the relevant link by the RA algorithm, are simply copied in the related descriptor.
This means that there is no need to raise additional interrupts, besides those generated at the end of frame transmissions and receptions.
It is worth remarking that neither RA algorithms, nor mechanisms like the TX series, are directly part of IEEE 802.11 \cite{IEEE802-11-2021} specifications.

\subsection{Combined Steering (TXreq+TXOP)}
A split mechanism very similar to RA, 
we named Combined Retransmission and Steering (CRS), 
can be used to provide highly-configurable, yet easily implementable \mbox{U-MAC} operation.
What is needed is a way to early encode the information to support \mbox{L-MAC} selection in the packet descriptor, so that optimal late steering decision can be made by the network adapter.
Some preliminary schemes will be presented below, which are part of our ongoing activities on the subject.
In the following, we will assume that every AC has its own transmission queue, and packets are not allowed to move between different queues (which is how real adapters customarily work).
However, strategies for jointly managing both ACs and \mbox{L-MACs} are in theory possible.

The simplest approach is to include a bitmap in every packet descriptor that specifies the channels on which it can be sent (having all bits set to zero is not allowed).
Assuming that an MLD can have at most four \mbox{L-MACs} (typically they have two or three), $4$ bits are enough.
When one of the DCUs gains a TXOP (on one of the \mbox{L-MACs}), its queue is orderly scanned looking for packets that can be sent (according to their bitmap).
If a single bit was set by the optimization algorithm executing on the host, then transmission can take place only on the specified channel, as in the early approaches described above.
Conversely, setting all bits means that all channels can be exploited, which closely resembles the late FIFO policy.

An enhanced version of the above approach foresees a distinct bitmap for every TX series.
In this case, two bytes ($4 \times 4$ bits) are enough to encode all possible options.
This seemingly minor modification has a tangible impact on the policies that can be implemented.
For example, it is possible to state that a certain number of attempts must be initially performed on a given channel, selected according to specific rules,
subsequently enlarging (in a controlled way)
the set of channels if packet transmission repeatedly incurs in failures.
For example, high-capacity background traffic exchanges can be confined to the $\unit[5]{GHz}$ band,
while the $\unit[6]{GHz}$ band is reserved to time-aware applications (e.g., multimedia and closed-loop control).
If none of them succeeds, the following attempts can be allowed to exploit TXOPs on other channels as well, to prevent packets from being dropped by the MAC layer when temporary communication outages are experienced on the intended channel.
As a last resort, the final series of attempts could employ, e.g., also the slower $\unit[2.4]{GHz}$ band, otherwise reserved for communication with legacy devices.

The exact way the host configures link usage in every packet descriptor can rely on both statistics about channels collected from \mbox{L-MACs} (average latency, frame loss ratio, and even latency percentiles) and the characteristics of the set of data flows currently mapped on any given AC (average and peak traffic, generation pattern, deadlines, desired reliability), and is the subject of future work,
along with performance evaluation of the proposed mechanisms.

When digital twins \cite{2022-IEEE-CSM, scanzio2024multilink} are exploited for optimizing the overall behavior of an industrial communication system (customarily made up of a number of infrastructure \mbox{Wi-Fi} networks interconnected through an Ethernet backbone), the amount of information potentially available to the STAs to support packet steering can be quite large.
Moreover, the exact nature of the different pieces of information that can be used to this purpose may vary consistently among setups, which demands for highly-flexible software implementations.
The point is that, only a limited number of simple rules can be realistically managed directly by the adapter.
Conversely, optimization algorithms running on the host can be arbitrarily complex, to the point that they can even rely on ML.

\section{Conclusions}
\label{sec:concl}
MLO in \mbox{Wi-Fi} 7 is expected to make communication more reliable and deterministic, by achieving lower transmission latency and jitters.
Basically, it permits the network adapter to select which channel to use on a per-packet-stream basis, choosing among a small set of active links (for which association to the related AP has been done in advance).
It is clear that the more sophisticate the selection strategy, the more effective optimization could be.
On the downside, complex and flexible procedures, needed to get the most out of the already crowded frequency ranges, can be hardly carried out by the sole network adapter, which has limited computation resources.

In this paper, some practical mechanisms for packet steering are described that enable inexpensive implementations.
The basic idea is derived from the hardware support to TX series on which rate adaptation algorithms rely: 
part of the information used for steering, e.g., related to the requirements and constraints of distributed applications, is evaluated in software by the host and stored in the packet descriptor when it is enqueued.
When gaining a transmission opportunity on some channel, the network adapter (implemented as a MLD) will then use the information in the descriptor to decide which packets will be actually sent.
Possibly, additional information available at that time about, e.g., the current state of the channel on which the TXOP was obtained, could be employed.

For the sake of truth, as far as we know very limited information is publicly available at this time about the internal architecture of \mbox{Wi-Fi} 7 chipsets, and so we were not able to perform a comparison with commercial solutions yet.
Moreover, the firmware of commercial \mbox{Wi-Fi} equipment is often not publicly available, which makes it difficult for the scientific community to propose new solutions.
As soon as we will manage to grasp some relevant data sheets, we will refine our analysis and redefine our proposal.

\bibliographystyle{IEEEtran}
\bibliography{bibliography}

\end{document}